# Behaviour of a Bucky-ball under Internal and External Pressures

Narinder Kaur[1, 2], Shuchi Gupta[1, 3], Keya Dharamvir[1] and V. K. Jindal[1]

[1]Department of Physics, Panjab University, Chandigarh, India

[2] Chandigarh College of Engineering and Technology, Chandigarh, India.

[3] University Institute of Engineering and Technology, Panjab University, Chandigarh, India.

*We study the behaviour of the $C_{60}$ molecule under very high internal or external pressure using Tersoff as well as Brenner potentials. As a result, we estimate the critical internal and external pressures that lead to its instability. We also calculate stretching force constant and bulk modulus of this molecule at several pressures under which the molecule remains stable. The values of these estimated here at zero pressure agree closely with those obtained in earlier calculations. We also observe that at high pressures, a finite value of parameter $\lambda_3$ of Tersoff potential gives physically acceptable results in contrast to its value zero, which is usually taken for the carbon systems.*

## 1 Introduction

The $C_{60}$ molecule, also called bucky-ball, is quite resistant to high speed collisions [1]. In a bucky ball, the atoms are all interconnected with each other through $sp^2$ bonding, thus resulting in exceptional tensile strength. In fact, the bucky ball can withstand slamming into a stainless steel plate at 15,000 mph, merely bouncing back, unharmed. When compressed to 70 percent of its original volume, the bucky ball is expected to become more than twice as hard as diamond [1].

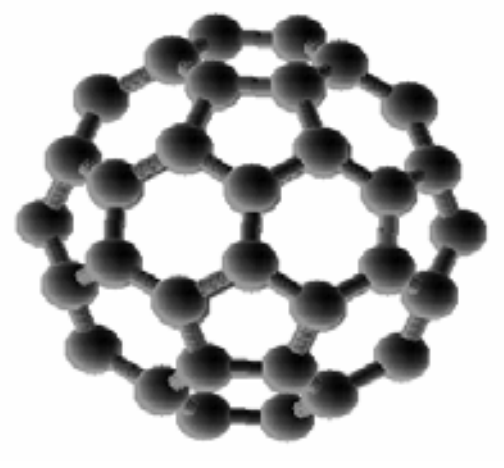

**Fig 1**: A $C_{60}$ molecule

Apart from its hardness, the important fact is that for nanotechnology, useful dopant atoms can be placed inside the hollow fullerene ball. This could be useful for a number of practical applications, the most notable being in the field of medicine. Drugs could be administered molecularly, or more importantly, individual radioactive molecules could be contained within the bucky ball for specific treatment of cancer. This will be more effective as compared to radiotherapy, since using bucky balls, one will be able to bombard the patient with low level (yet relatively large effective quantities) of radiation [2] and hence less of side effects. In order to utilize such properties and the strength of this molecule, it is of interest to study its stability under internal and external pressure. Therefore, in this paper, we made an attempt to study the stability of this molecule based on its binding strength. We have used the Tersoff [3, 4] as well as Brenner potential [5] for the intramolecular interactions between the carbon atoms of a bucky ball and compared the two**.**

There are a number of theoretical and experimental studies available for the phonon modes in $C_{60}$ molecule. Experimental data from Raman scattering [6], Infrared [6, 7] and neutron inelastic measurements [8] provide an overview of the vibrational modes of the $C_{60}$ molecule. Similarly, several calculations of the vibrational frequencies of the $C_{60}$ molecule, using various classical and quantum mechanical theories [9-12], have been performed. In addition, a number of force constant models have been used to calculate the phonon frequencies [13, 14]. For calculating the phonon frequencies, they had first fitted the Raman data to obtain the force constants and then various phonon frequencies had been calculated using these force constants. In order to show the validity of the potential model used by us, we have obtained the stretching force constant of a $C_{60}$ molecule, which would shown to be in good agreement with the theoretical work done by Jishi et.al. [13] and hence with the experimental observations.

In order to utilize the hardness of a $C_{60}$ molecule as molecular bearings etc., its Bulk Modulus needs to be evaluated. Ruoff and Ruoff [15] estimated the bulk modulus of $C_{60}$ from simple elasticity theory as 843 GPa and, using the Tight Binding method, Woo et.al. [16] found its value to be 717 GPa. These values have been calculated around zero pressure. In this paper, Bulk Modulus for the $C_{60}$ molecule has been estimated for higher pressures as well.

In section 2 we present the theoretical model used to obtain equilibrium structure of a bucky ball. The numerical method and results have been presented in section 3. Section 4 contains discussions and conclusions.

## 2 Theoretical model

We have used a theoretical model in which the interaction between bonded carbon atoms is governed by (i) Tersoff potential [3, 4] and (ii) Brenner potential [5]. These potentials have been extensively used to interpret properties of several carbon based systems like carbon nanotubes [17], graphite [3-5], diamond [3-5] and fullerenes [18]. These potentials are also suitable for silicon and hydrocarbons [4, 5]. These potentials are able to distinguish among different carbon environments, fourfold $sp^3$ bond as well as threefold $sp^2$ bond. A comparison of results of structure and bulk modulus under high pressures from these two potentials helps us explore their applicability in high pressure regime.

### 2.1 Tersoff potential

The form of this potential is expressed as potential energy between any two carbon atoms on $C_{60}$, say $i$ and $j$, separated by a distance $r_{ij}$ as

$$V_{ij} = f_c(r_{ij})(Ae^{-\lambda_1 r_{ij}} - b_{ij}Be^{-\lambda_2 r_{ij}}) \tag{1}$$

$f_c(r)$ is a function used to smooth the cutoff distance taken as 2.1Å. It varies from 1 to 0 in sine form between 1.8 Å and 2.1 Å [3]. The state of the bonding is expressed through the term $b_{ij}$ as the function of angle between bond *i*-*j* and each neighboring bond *i*-*k* (see Fig 2).

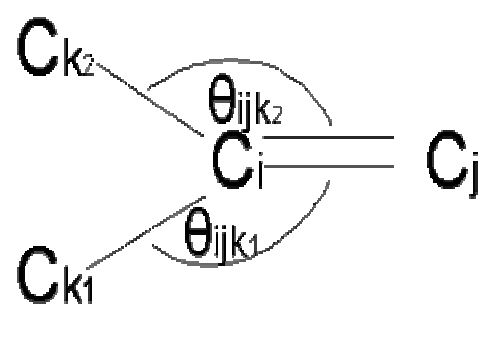

**Fig 2**: A set of four neighbouring carbon atoms

The parameters used in this potential have already been given by Tersoff [4]. However some of them have been modified by us (as discussed in section 3.1, para 2) and have been tabulated in Table I.

**Table I:** The original and modified parameters of the Tersoff potential.

| Tersoff Parameters | Original [4] | Modified |
|---|---|---|
| $A$ (eV) | 1393.6 | 1380.0 |
| $B$ (eV) | 346.7 | 349.491 |
| $\lambda_1$ ( Å$^{-1}$) | 3.4879 | 3.5679 |
| $\lambda_2$ ( Å$^{-1}$) | 2.2119 | 2.2564 |
| $\lambda_3$ ( Å$^{-1}$) | 0 | 0, 2.2564 |

## 2.2 Brenner potential

The potential energy between any two carbon atoms on $C_{60}$, say *i* and *j*, separated by a distance $r_{ij}$ is given as

$$V_{ij} = V_R(r_{ij}) - B_{ij} V_A(r_{ij}), \tag{2}$$

where,

$$V_R(r_{ij}) = f(r_{ij}) \frac{D_e}{S-1} \exp\left[-\beta\sqrt{2S}(r - R_e)\right], \tag{3}$$

$$V_A(r_{ij}) = f(r_{ij}) \frac{D_e S}{S-1} \exp\left[-\beta\sqrt{\frac{2}{S}}(r - R_e)\right], \tag{4}$$

$$f(r_{ij}) = \begin{cases} 1, \; r < R_1 \\ \frac{1}{2}\left[1 + \cos\left(\frac{(r_{ij} - R_1)\pi}{(R_2 - R_1)}\right)\right], R_1 < r_{ij} < R_2 \\ 0, \; r_{ij} > R_2 \end{cases} \tag{5}$$

$V_R$ and $V_A$ are the repulsive and attractive potential energy terms, respectively, which together form a sort of modified Morse potential. The screening function $f(r_{ij})$ restricts the interaction to nearest neighbors, as defined by the values for $R_1$ and $R_2$. In addition, the Brenner potential takes bonding topology into account with the empirical bond order function $B_{ij}$, whose exact form has been taken from Ref.[5]. Brenner has defined two sets of parameters, henceforth defined as set-1 and set-2, for this potential for carbon systems [5], which have been presented in Tables II and III.

**Table II:** Set–1 parameters of the Brenner potential.

| $D_e$ | $S$ | $\beta$ | $R_e$ | $R_1$ | $R_2$ | $\delta$ | $a_0$ | $c_0$ | $d_0$ |
|---|---|---|---|---|---|---|---|---|---|
| 6.325 eV | 1.29 | 1.5 Å$^{-1}$ | 1.315 Å | 1.7 Å | 2.0 Å | 0.80469 | 0.011304 | 19 | 2.5 |

**Table III:** Set- 2 parameters of the Brenner potential.

| $D_e$ | $S$ | $\beta$ | $R_e$ | $R_1$ | $R_2$ | $\delta$ | $a_0$ | $c_0$ | $d_0$ |
|---|---|---|---|---|---|---|---|---|---|
| 6.0eV | 1.22 | 2.1 Å$^{-1}$ | 1.39Å | 1.7 Å | 2.0 Å | 0.5 | 0.00020813 | 330 | 3.5 |

Using these potentials, composite energy of all the atoms of the system, given by $E$ is written as

$$E = \sum_{ij} V_{ij} \quad , \tag{6}$$

where the sum over *i* and *j* in Eq.6 includes all the 60 atoms in the $C_{60}$ molecule.

## 3 Numerical method and Results

We discuss here the details of numerical methodology and results. We give the essential ingredients and then describe effects of pressure on the molecule.

### 3.1 Structure and Potential parameters

The structure of $C_{60}$ is a truncated icosahedron, which resembles a round soccer ball of the type made of hexagons and pentagons, with a carbon atom at the corners of each hexagon and a bond along each edge. Two types of bond lengths determine the coordinates of 60 carbon atoms in $C_{60}$ molecule. Single bond $b_1$, also called the 6:5 ring bond, joins a hexagon and a pentagon. The double bond $b_2$, also called the 6:6 ring bond, joins two hexagons and is shorter. These have been measured using nuclear magnetic resonance, and are found to be having lengths 1.46Å and 1.40Å [2] respectively.

By using the parameters given by Tersoff, the structure was allowed to minimize using the potential model as given in the earlier section 2.1. In this way, $b_1$, $b_2$ and bond angles were varied to obtain minimum energy configuration. By doing this, at zero pressure, $b_1$ and $b_2$ were obtained to be 1.46Å and 1.42Å with binding energy 6.72eV/atom as given in Table IV. In order to reproduce the bond lengths and the binding energy of $C_{60}$ molecule in closer agreement with the experimental results, the potential parameters given by Tersoff [4] had to be modified. It was found that Tersoff parameters $A$, $B$, $\lambda_1$ and $\lambda_2$ were more sensitive parameters to get appropriate binding energy and bond lengths so only these were modified. Tersoff has taken $\lambda_3$ equal to zero, however in order to see its effect, we have also used a finite value of $\lambda_3$. It has been found that the effect of change in $\lambda_3$ becomes evident only at high pressures. In Table I, we have tabulated the modified as well as the original potential Parameters [4]. The new bond lengths and energies have been given in Table IV.

**Table IV**: Comparison between the calculated and experimental Binding energy and bond lengths of a $C_{60}$ molecule with original and modified parameters.

| | Calculated | | | | Experimental [2] |
|---|---|---|---|---|---|
| | Tersoff potential | | Brenner potential | | |
| | original [4] parameters | parameters modified by us | set-1 [5] parameters | set-2 [5] parameters | |
| Binding energy (eV/atom) | -6.73 | -7.17 | -7.04 | -6.99 | -7.04 |
| Bond lengths (Å) ($b_1, b_2$) | (1.46,1.42) | (1.45,1.41) | (1.45,1.42) | (1.48,1.45) | (1.45,1.40) |

Similarly, energy minimization has been done with Brenner potential using both set-1 and set-2 parameters and the binding energy/atom and bond lengths so obtained for the minimized structures have been tabulated in Table IV.

### 3.2 Pressure effects

Application of pressure $P$ on the molecule decreases its volume by $\Delta V$ and increases the binding energy $E$ of the molecule by $P\Delta V$ in accordance with the equation

$$E(P) = E(0) + P\Delta V \tag{7}$$

To compress or dilate the molecule we multiply each coordinate of 60 atoms by a constant factor $C \langle 1$ for compression and $C \rangle 1$ for dilation. Each $C$ value determines a +ve (external) or a –ve (internal) pressure. By changing $C$, we get a new diameter d, (hence a new volume $(V)$) and a new binding energy ($E$) as shown in Fig 3. Thus we get $E$ as a function of $V$. Change in volume represents some application of pressure on the ball. This pressure has been obtained by calculating the first derivative of the molecular energy w.r.t its volume. Pressures thus obtained corresponding to various diameters of interest are shown in Fig 4. From fig 3 and 4 we get $E(P)$.

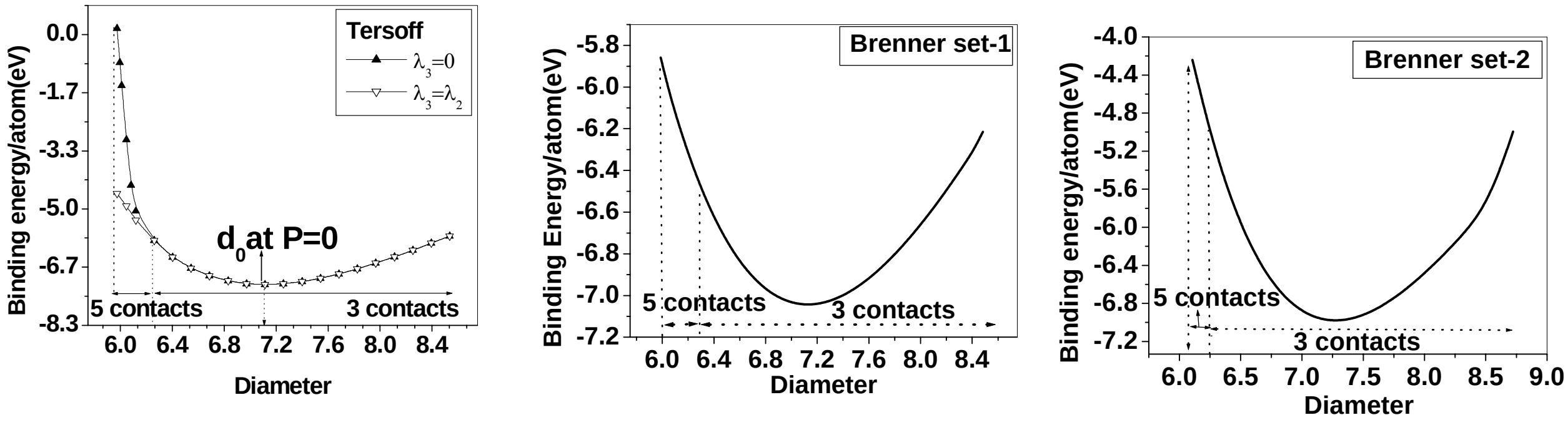

(a) (b) (c)

**Fig 3:** Binding energy, E, of a relaxed $C_{60}$ molecule at different diameters under Tersoff (a), Brenner set-1 (b) and set-2 (c) potentials.

We have made the assumption that the shape of the molecule does not change with pressure. This must be true when one deforms the regular $C_{60}$ hydrostatically. Theoretically, this can easily be done by first converting Cartesian coordinates (x,y,z) of 60 atoms into polar coordinates $(r,\theta,\phi)$ and then minimizing the structure allowing only $\theta$ and $\phi$ to change at a fixed radius $r$ of $C_{60}$ molecule.

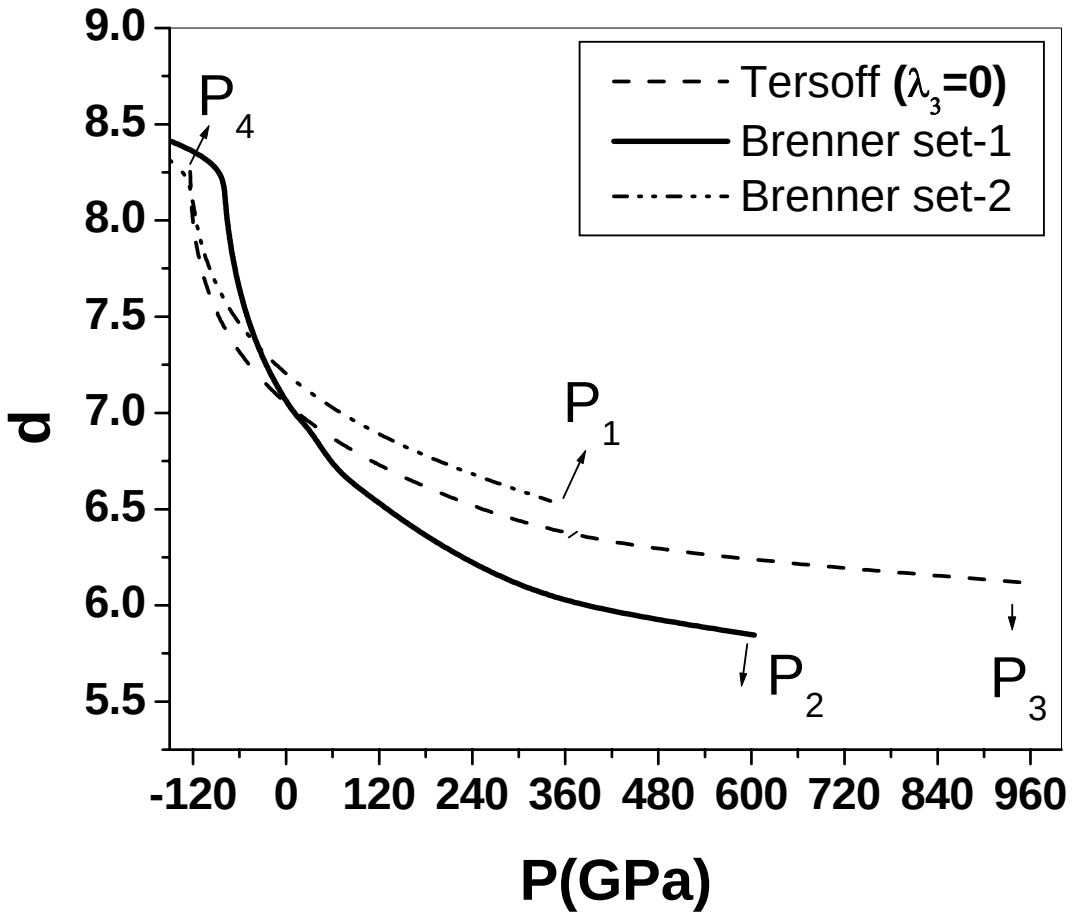

**Fig 4:** Pressure required to achieve a particular diameter of the ball for the three potential models. $P_1$, $P_2$ and $P_3$ are the critical pressures at which $C_{60}$ molecule becomes unstable. $P_4$ represents the breaking point of the molecule, under Tersoff potential.

At zero pressure, each carbon atom in $C_{60}$ has coordination number $N$ as 3 i.e. there are three nearest neighbors within the range of the potential and the bonds with these atoms are shown bold in Fig 5a. At some pressure, $N$ changes from 3 to 5 as two more atoms come inside the range of the potential as shown bold in Fig. 5b. The value of this pressure is dictated by the potential model. The variation in the coordination number with increase in diameter (pressure) using Tersoff potential is shown in Table V.

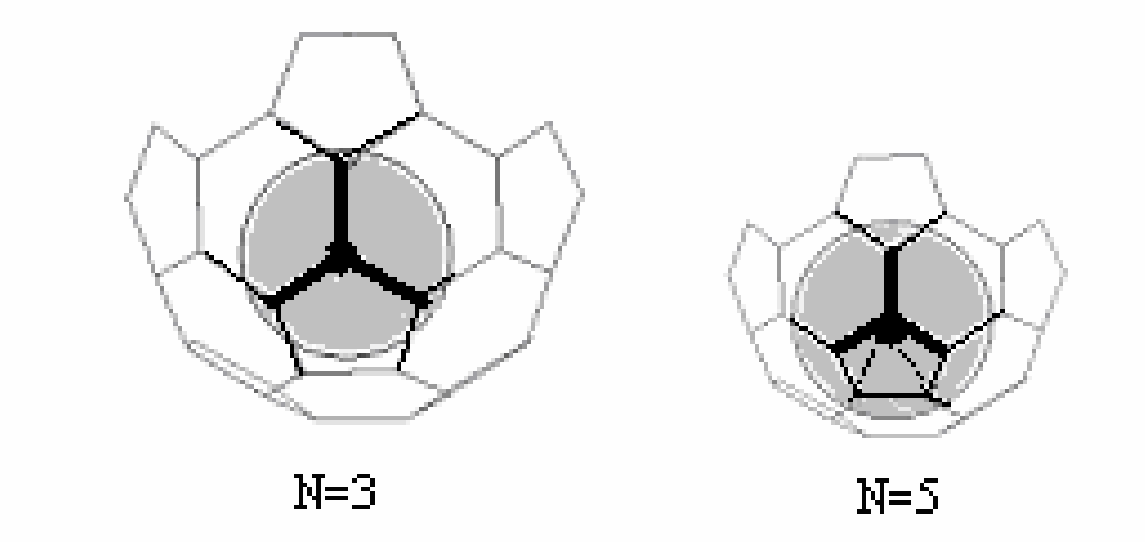

(a) (b)

**Fig 5:** Number of carbon atoms within the range of short range potentials at different pressures.

**Table V:** Co-ordination number of each carbon atom for different diameter (using Tersoff potential)

| $D$(Å) | 6.4 | 6.0 | 5.3 | 4.2 | 4.0 | 3.9 |
|---|---|---|---|---|---|---|
| $N$ | 3 | 5 | 9 | 11 | 15 | 17 |

An inspection of Fig 3a reveals that at a diameter $d_1$ = 6.367 Å or less, the rise in energy is faster for $\lambda_3 = 0$ as compared to that for $\lambda_3 = \lambda_2$ where $\lambda_2 = 2.2564$. For a higher value of $\lambda_3$, during compression the value of bond order term $b_{ij}$ decreases quickly, which in turn appreciably decreases the attractive part of the potential (see equation 1). This explains the sudden increase in $E$ for higher $\lambda_3$.

**a) Critical diameters:**

The molecule can withstand internal and external pressure upto a certain extent. To have a knowledge of these limits, a plot between binding energy/atom, $E(P)$ and diameter of the ball, $d$ as in fig. 3 is used. We have been able to minimize the energy of the $C_{60}$ molecule having diameter within certain range, depending upon the potential used, but outside this range minimum energy configuration was not obtained even after a very large number of cycles of iteration. These critical diameters and corresponding pressures have been tabulated in Table VI.

For diameter less than $d_1$, the repulsive potential becomes so large that the molecule becomes unstable and for diameter greater than $d_2$, some of the bonds get broken and the molecule does not remain a closed caged structure. The structures at these critical diameters have been shown in fig. 6 and fig. 7 shows the relative volume of the molecule at different pressures. Maximum dilation or compression in terms of volume has been estimated here under the three potentials. These values under Tersoff, Brenner set-1 and Brenner set-2 potentials are (225%, 59%), (191%, 49%) and (173%, 59%) respectively. Critical diameters under the three potentials are presented in Table VI.

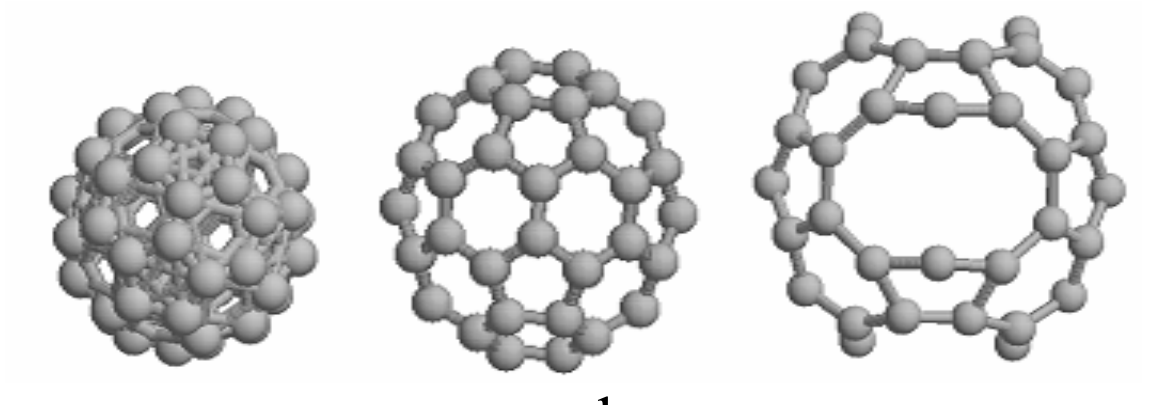

**Fig 6:** Molecular structure at critical diameters. The structures 'a', 'b' and 'c' correspond to extreme external, zero and extreme internal pressures respectively.

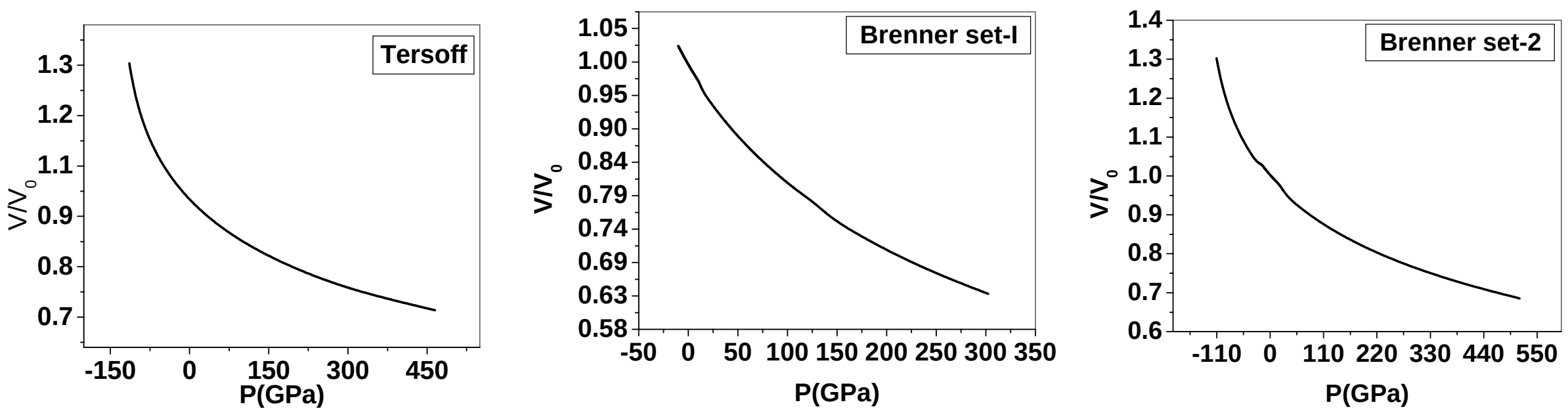

**Fig 7:** Calculated P-V curve for three models.

**Table VI**: Critical diameters and pressures under different Potentials.

| Potential | $d_1$ (Å) | $P_1$(GPa) | $d_0$ (Å) P=0 | $d_2$ (Å) | $P_2$(GPa) |
|---|---|---|---|---|---|
| Tersoff | 5.98 | 922 | 7.11 | 9.32 | -134 |
| Brenner set –1 | 5.63 | 535 | 7.13 | 8.84 | -116 |
| Brenner set –2 | 6.11 | 322 | 7.27 | 8.72 | -131 |

**b) Force constant**

Due to the application of pressure the bond length decreases, say, by $x$ and bond energy increases by $\partial E$, as shown in fig 8 and related through equation 8.

$$\partial E = \frac{1}{2} k x^2 \qquad (8)$$

$$k = \frac{\partial^2 E}{\partial x^2} \qquad (9)$$

Double derivative of the binding energy of the molecule with respect to its bond length as in equation 9, give the value of force constant. In Table VII, we compare the value of bond stretching force constant with other similar work.

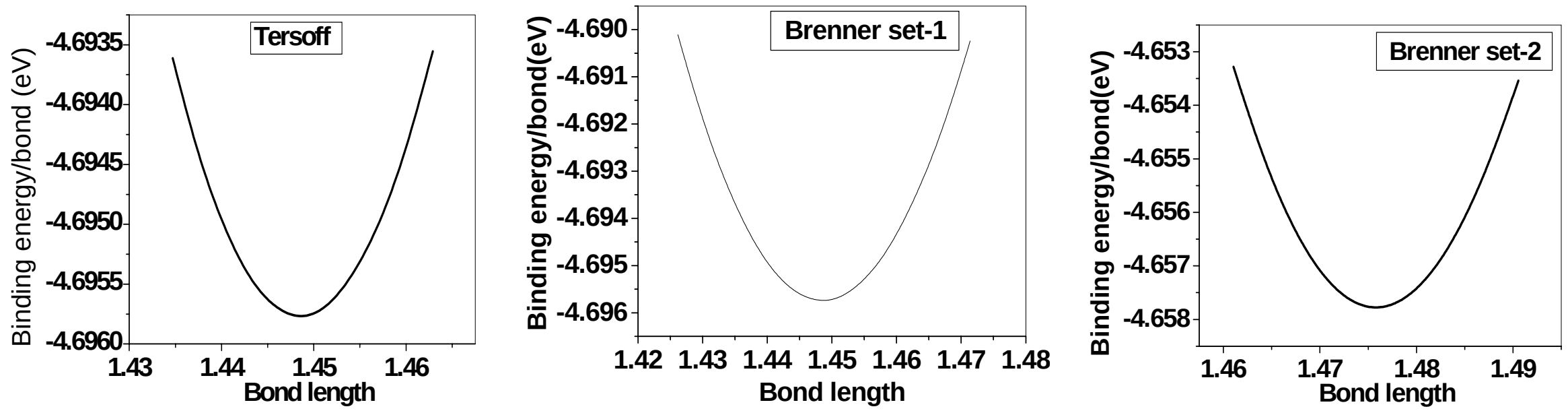

a b c

**Fig. 8:** Binding Energy/bond of $C_{60}$ molecule with varying single bond lengths.

**Table VII:** Force constants of bond stretching of a $C_{60}$ molecule (mdyne/Å)

| Tersoff | Brenner set-1 | Brenner set-2 | Jishi.et.al. [13] | Ruoff and Ruoff [15] | graphite [2] | Feldman et. al. [14] | Cylvin et.al. [19] |
|---|---|---|---|---|---|---|---|
| 5.6 | 3.55 | 6.51 | 4.0 | 6.62 | 3.5 | 4.4 | 4.7 |

### c) Bulk modulus

An application of a hydrostatic pressure $P$ alters the total binding energy $E$ of the molecule such that

$$\partial E = -P\Delta V \qquad (10)$$

where $\Delta V$ is the change in volume and $\partial E$ is the increase in the binding energy.

$$-\frac{\partial E}{\partial V} = P \qquad (11)$$

Bulk modulus of the molecule indicates its hardness at different pressures and has been calculated using the equation

$$B = -V_0 \frac{\partial P}{\partial V} \qquad (12)$$

$$B = V_0 \frac{\partial^2 E}{\partial V^2} \tag{13}$$

Second derivative of the binding energy of the molecule with respect to its volume gives us the bulk modulus and is shown in Fig 9. Ruoff and Ruoff [15] have calculated the bulk modulus of this molecule, using force constant for bond stretching using the data presented in Table VII, as 843Gpa. Woo et. al. [16] have also calculated bulk modulus (717GPa) by studying the dynamics of the molecule using Tight Binding method. The value of the bulk modulus around zero pressure calculated under Tersoff, Brenner set-1 and set-2 comes out to be 674GPa, 370GPa and 694GPa respectively. A comparison of various calculations has also been made in Table VIII.

**TableVIII**: Bulk modulus, average bond length and radius of a $C_{60}$ molecule according to various calculations.

| Reference | Radius (Å) | Av. bond length (Å) | Bulk modulus (GPa) |
|---|---|---|---|
| Ruoff.et.al | 3.52 | 1.43 | 843 |
| Woo.at.al | 3.57 | 1.43 | 717 |
| Tersoff | 3.56 | 1.43 | 674 |
| Brenner set-1 | 3.56 | 1.42 | 370 |
| Brenner set-2 | 3.64 | 1.46 | 694 |

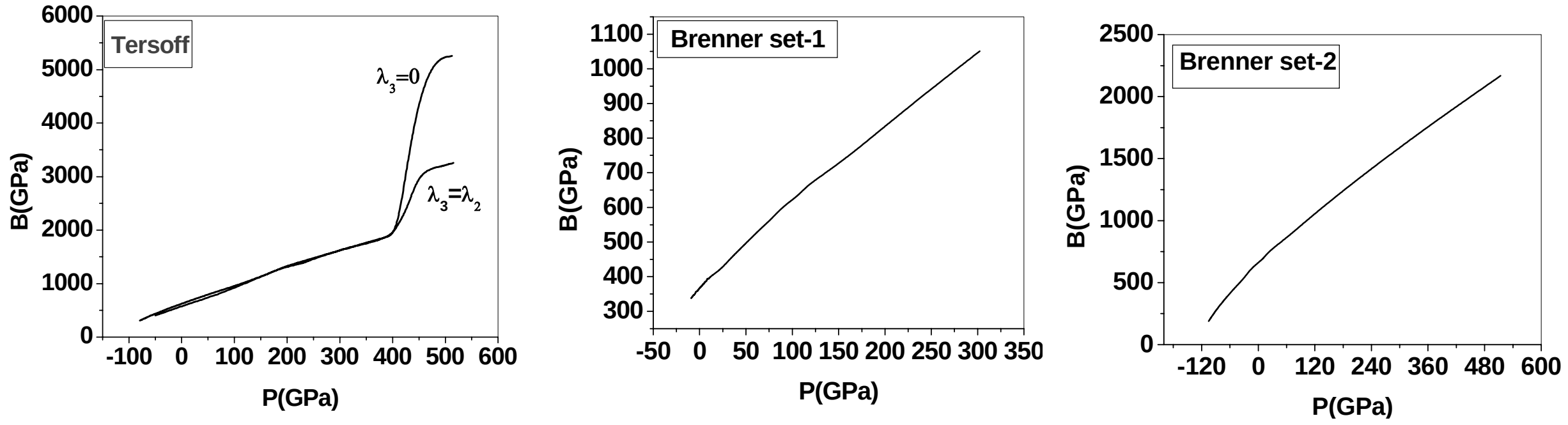

**a** **b** **c**

**Fig 9:** Variation of bulk modulus with pressure

Comparing the plots in fig we find (see fig. 9a) that the hardness of the molecule increases suddenly under Tersoff potential at a pressure above 400GPa, when the volume compression is 73%. Such abruptness is missing in figs. 9 b and c.

## 4 Discussion

In this paper, we attempt to calculate pressure effects on a $C_{60}$ molecule using Tersoff as well as the Brenner potential for the inter-atomic interactions. With this pressure study, we have been able to calculate the critical diameters, stretching force constant, and Bulk Modulus of a $C_{60}$ molecule at wide ranging pressure values.

Although Tersoff potential reproduces the experimental results around zero pressure but at high pressures, as the coordination number increases, the bulk modulus show discontinuity in its value. This effect gets marginalized by altering the parameter $\lambda_3$ to a non zero value which has usually been taken to be zero for carbon systems. We have recalculated the results by taking $\lambda_3 = \lambda_2$ which has already been used to explain the silicon system successfully and fig.9a shows that the discontinuity in the bulk modulus gets diminished by choosing this value. There is a scope for further improvement in the value of this parameter. Usage of Brenner potential does not show such discontinuity.

We also explore the stability of the molecule, under extreme external and internal pressure. Maximum compression or dilation in terms of volume of $C_{60}$ molecule has been estimated here, by using three potential models. The critical pressures corresponding to maximum compression or dilation are about 922Gpa and -134Gpa respectively with Tersoff potential, 535Gpa and -116GPa respectively with Brenner set-1 potential, and 322Gpa and -131Gpa respectively with Brenner set-2 potential. The reliability of these potentials at these extreme pressures is the relevant question here. One has to resort to ab-initio calculations to get exact values, though the estimates provided here using Brenner set-1, can be considered to be quite reliable as this has provided good estimate of binding energy and bond lengths of $C_{60}$ molecule.

As shown in Table IV, bond lengths and binding energy calculated by using Brenner set-1 are in close agreement with the experimental value as compared to Brenner set-2. However, the force constant and Bulk modulus calculated using Brenner set-2 are in better agreement with the literature values as shown in Tables VII and VIII. In Ref. [5] Brenner also noted the same. Although modified Brenner potential [20] could have been used, it was considered unnecessary in the present calculation as our aim was to study the pressure effects on the molecular structure.

Though the Bulk modulus calculated by using Tersoff and Brenner set-2 potential, around zero pressure are approximately the same, but at high pressure (see fig 9) there is a sudden jump in the value of bulk modulus under Tersoff potential in contrast to the results obtained from Brenner potential. It is quite unexpected that a molecule becomes suddenly so hard. Therefore Tersoff potential is not suitable above approximately 400 GPa. However the molecule's behaviour at high pressures may become physically acceptable by choosing appropriate value of parameter $\lambda_3$ if we insist on using this potential.

We believe that the estimates of critical positive and negative pressures can be useful for planning practical applications related to either high pressure absorption or doping of atoms or molecules in the cage.

The results of bulk moduli provide enough motivation for further measurements on this molecule under high pressures.

**Acknowledgements**

VKJ wishes to acknowledge financial support from TBRL, DRDO in the form of a research project.

**References**

1. Curl, Robert F. and Richard E. Smalley. "Fullerenes" Scientific American 265, 54 (October 1991), .
2. M. S. Dresselhaus, G. Dresselhaus and P. C. Eklund, Science of Fullerenes and Carbon Nanotubes, Academic Press, California 1996.
3. J. Tersoff, Phys. Rev. B 37, 6991 (1988).
4. J. Tersoff, Phys. Rev. Lett. 61, 2879 (1988).
5. D. W. Brenner Phy. Rev. B 42, 9458 (1990).
6. D. S. Bethune, G. Meijer, W. C. Tang, H. J. Rosen, W. G. Golden, H. Seki, C. A. Brown, and M.S. Derries, Chem. Phys. Lett. 179,181 (1991).
7. D. S. Bethune, G. Meijer, W. C. Tang, H. J. Rosen, Chem. Phys. Lett. 174, 219 (1990).
8. R. L. Cappelletti, J. R. D. Copley, W. Kamitakihara, Fang Li, J. S. Lannin and D. Ramage, Phys. Rev. Lett. 66, 3261 (1991).
9. R. E. Stanton and M. D. Newton, J. Phys. Chem. 92, 2141(1988).
10. F. Negri, G. Orlandi and F. Zerbetto, Chem. Phys .Lett. 144, 31 (1988).
11. B. P. Feuston, W. Andreoni, M. Parrinello and E. Clementi, Phys.Rev.B 44, 4056 (1991).
12. G. B. Adams, J. B. Page, O. F. Sankey, K. Sinha, J. Menendez and D. R. Huffman, Phys. Rev. B 44, 4052(1991).
13. R. A. Jishi, R. M. Mirie, M. S. Dresselhaus, Phys. Rev. B 45, 13685 (1992).

14. J. L. Feldman, J. Q. Broughton, L. L. Boyer, D. E. Reich and M. D. Kluge, Phys. Rev. B 46, 12731 (1992).
15. R. S. Ruoff, A. L. Ruoff, Appl. Phys. Lett. 59 (13), 23 (1991).
16. S. J. Woo, S. H. Lee, Eunja Kim, K. H. Lee, Young Hee Lee, S. Y. Hwang and Cheol Jeon Phys. Lett. A 162 (1992).
17. S. Gupta, K. Dharamvir and V. K. Jindal, Phy. Rev. B 72, 165428 (2005).
18. D. H. Robertson, D. W. Brenner and C.T.White, J. Phys. Chem. 1999, 15721 (1994).
19. S. J. Cyvin, E. Brendsdal, B. N. Cyvin and J. Brunvoll, Chem. Phys. Lett. 174, 219 (1990).
20. D.W. Brenner, O.A. Shenderova, J. A. Harrison, S.J. Stuart, B. Ni and S. B. Sinnott, J. Phys. Condensed Matter 14, 783 (2002).